\documentclass{article}
\usepackage{spconf,amsmath,graphicx,hyperref}
\usepackage{amsmath,amssymb,amsfonts}
\usepackage{graphicx}
\usepackage{cite}
\usepackage{url}
\usepackage{booktabs}
\usepackage{algorithm}
\usepackage{algorithmicx}
\usepackage{multirow}
\usepackage{xcolor}
\usepackage{subcaption}
\usepackage{siunitx}
\usepackage{booktabs}
\usepackage{algpseudocode}

\usepackage{xcolor} 
\makeatletter
\def\ALG@numberfreq{1}
\algrenewcommand\alglinenumber[1]{%
  \scriptsize\textcolor{gray}{\ifnum#1>0 \ifnum#1<10 0\fi #1:\fi}%
}
\makeatother

\title{Membership Inference Attack Against Music Diffusion Models via Generative Manifold Perturbation}
\name{Yuxuan Liu, Peihong Zhang, Rui Sang, Zhixin Li, Yizhou Tan, Yiqiang Cai, Shengchen Li\thanks{The project demo is available at: https://github.com/kaslim/LSA-Probe}}
\address{Xi'an Jiaotong-Liverpool University, Suzhou, China}
\begin{document}
\ninept
\maketitle

\begin{abstract}
Membership inference attacks (MIAs) test whether a specific audio clip was used to train a model, making them a key tool for auditing generative music models for copyright compliance. However, loss-based signals (e.g., reconstruction error) are weakly aligned with human perception in practice, yielding poor separability at the low false-positive rates (FPRs) required for forensics. We propose the Latent Stability Adversarial Probe (LSA-Probe), a white-box method that measures a geometric property of the reverse diffusion: the minimal time-normalized perturbation budget needed to cross a fixed perceptual degradation threshold at an intermediate diffusion state. We show that training members, residing in more stable regions, exhibit a significantly higher degradation cost. With matched compute and a fixed threshold ($\tau{ = }\mathrm{P95}$), LSA-Probe improves TPR@1\% FPR by 3–8 percentage points over the best baseline across DiffWave and MusicLDM. These results indicate that local generative stability provides a reliable membership signal for audio diffusion models.

\end{abstract}


\section{Introduction}
Diffusion models are revolutionizing music generation, synthesizing high-fidelity audio that achieves a high degree of perceptual realism~\cite{chen2024musicldm,chowdhury2024melfusion}. However, the remarkable performance of these models is predicated on their training over vast datasets, often of uncertain provenance and consent~\cite{dornis2024generative}. This dependency creates a significant risk of copyright infringement and privacy violations for artists and rights-holders. Consequently, there is an urgent need for reliable auditing tools to determine whether a specific music clip was used to train a given model.

Membership Inference Attacks (MIAs) are designed to exploit a model's memorization of its training data to distinguish members  (training-set samples) from non-members (held-out samples)~\cite{shokri2017membership, tirumala2022memorization}. In the domain of computer vision, such attacks operate on a core assumption: member samples yield a lower reconstruction loss than non-member samples~\cite{hilprecht2019monte}. However, a growing number of recent studies indicates that this fundamental assumption is severely challenged when applied to audio domains. For instance, Rossi et al.~\cite{rossi2025membership} demonstrated that standard MIAs falter on sequence models because they fail to adequately model the complex internal correlations within the data. Kong et al. \cite{kong2024an} report that such endpoint, loss-only criteria are confounded and unstable for audio diffusion: they are sensitive to content complexity and timbral factors, exhibit high variance across noise seeds and timesteps. 



These findings have collectively catalyzed a paradigm shift in MIA research: a move away from relying on a single, static, endpoint signal (such as the reconstruction loss) towards exploiting more informative, dynamic, process-based signals~\cite{keskar2017on, liu2022membership}. This shift is motivated by an analogy to the well-established principle of "flat minima", where smoother regions of the loss landscape are strongly correlated with better generalization and robustness~\cite{keskar2017on}.  Matsumoto et al.~\cite{matsumoto2023membership} showed that members are more robust than non-members when adding the same adversarial perturbations.

Inspired by this, we propose the Latent Stability Adversarial Probe (LSA-Probe), a novel white-box framework that directly interrogates the geometric stability of the generative process to infer membership. Our key hypothesis is that a model learns a smoother, more stable generative mapping in the local vicinity of its training members. We operationalize this hypothesis by quantifying the adversarial cost: the minimum perturbation budget required to induce a fixed level of perceptual degradation when applied to an intermediate latent state of the reverse diffusion process. Because members reside in more stable regions of the generative manifold, the adversarial cost required to degrade them is significantly higher than for non-members, providing a robust signal for our attack.

The proposed method is evaluated on two types of music diffusion models: (i) a waveform Denoising Diffusion Probabilistic Model (DDPM)~\cite{ho2020denoising}, instantiated as DiffWave~\cite{kong2020diffwave}; and (ii) a Latent Diffusion Model (LDM), instantiated as MusicLDM~\cite{chen2024musicldm}—across two datasets: the MAESTRO v3~\cite{hawthorne2019maestro} piano performance corpus and the Free Music Archive (FMA) Large subset~\cite{defferrard2017fma}. Our experiments demonstrate a robust signal for membership that surpasses traditional baselines in the forensically crucial low-false-positive-rate regime. The primary contributions of this work are as follows:

\begin{enumerate}
\item To the best of our knowledge, this is the first systematic investigation of membership inference attacks for music diffusion models.
\item We propose the LSA-Probe, a white-box MIA that probes time-normalized latent stability along the reverse diffusion trajectory and scores samples by the adversarial cost needed to reach a fixed perceptual degradation.
\item We connect local generative stability to membership via a first-order analysis, yielding a practical score without requiring likelihoods or shadow models.
\end{enumerate}

\begin{figure*}[t]
  \centering
  \includegraphics[width=0.95\textwidth]{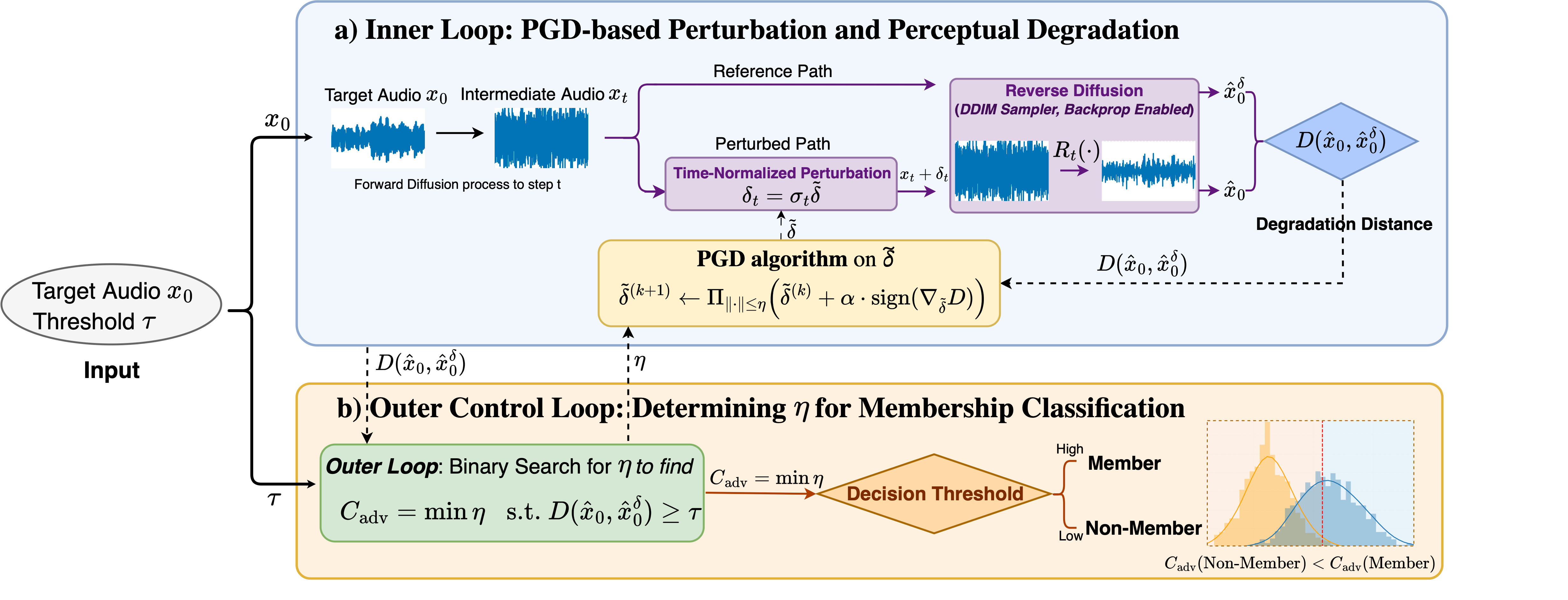}
  \caption{\textbf{LSA-Probe overview (two-loop procedure at timestep $t$).}
  \emph{(a) Outer control loop.} We search the budget $\eta$ by binary search to find the
  \emph{adversarial cost} $C_{\mathrm{adv}}(x_0;t,\tau)=\min\{\eta:\ D(\hat{x}_0,\hat{x}_0^{\delta})\ge\tau\}$,
  then use $C_{\mathrm{adv}}$ as the decision score (higher $\Rightarrow$ more likely \emph{member}); the dashed line indicates the decision threshold used for TPR@FPR analyses.
  \emph{(b) Inner loop (PGD on $\tilde\delta$).} Given a target waveform $x_0$, we form $x_t$
  by the forward process and inject a \emph{time-normalized} latent perturbation
  $\delta_t=\sigma_t\tilde\delta$ with $\sigma_t=\sqrt{1-\bar\alpha_t}$. We update
  $\tilde\delta$ for $K$ steps with projection onto the $\ell_p$ ball $\|\tilde\delta\|_p\le\eta$
  (sign step for $\ell_\infty$ or normalized gradient for $\ell_2$), while backpropagating
  through the deterministic reverse operator $R_t(\cdot;\theta)$ and the
  chosen differentiable distance $D$ (CDPAM, MR\mbox{-}STFT, log-mel MSE, or waveform MSE).
  The inner loop outputs the degradation $D(\hat{x}_0,\hat{x}_0^{\delta})$ for the current $\eta$,
  which the outer loop uses to adjust $\eta$. For each $(x_0,t)$ we fix the forward noise
  $\epsilon$ (seeded) to isolate the effect of $\delta_t$; restarts and momentum are omitted
  from the diagram for clarity.}
  \label{fig:overview}
\end{figure*}

\section{Proposed Method}
\label{sec:method}

We study a white-box, developer-side setting: the adversary has read-access to model parameters, the reverse sampler, and gradients w.r.t.\ inputs at intermediate diffusion states. The goal is to decide membership for a given waveform $x_0$ using only the audited model (no shadow models or training labels). An overview of our two-loop procedure (outer binary search for budget, inner PGD on a time-normalized perturbation) is shown in Fig.~\ref{fig:overview} and referenced throughout this section.

\subsection{Preliminaries: forward/noise parameterization and reverse operator}
Let $x_0\!\in\!\mathbb{R}^n$ denote a waveform. The forward noising at discrete step
$t\!\in\!\{1,\dots,T\}$ follows the usual DDPM parameterization
\begin{equation}
x_t \;=\; \sqrt{\bar\alpha_t}\,x_0 \;+\; \sqrt{1-\bar\alpha_t}\,\epsilon,\qquad 
\epsilon\!\sim\!\mathcal{N}(0,I),
\label{eq:forward}
\end{equation}
with schedule $\{\bar\alpha_t\}_{t=1}^T$. We adopt a deterministic DDIM-style reverse operator
with $\sigma_t{=}0$. Given a noise-predictor $\epsilon_\theta(\cdot,t)$, define the stepwise denoised
estimate
\begin{equation}
\hat{x}_0(x_t,t)
\;=\;
\frac{x_t - \sqrt{1-\bar\alpha_t}\,\epsilon_\theta(x_t,t)}{\sqrt{\bar\alpha_t}}.
\label{eq:x0hat}
\end{equation}
A single reverse step maps $x_t\!\mapsto\!x_{t-1}$ via
\begin{equation}
x_{t-1}
=
\sqrt{\bar\alpha_{t-1}}\,\hat{x}_0(x_t,t)
\;+\;
\sqrt{1-\bar\alpha_{t-1}}\,
\epsilon_\theta(x_t,t).
\label{eq:ddimstep}
\end{equation}
Composing \eqref{eq:ddimstep} from index $t$ down to $1$ yields a differentiable mapping
$R_t(\cdot;\theta):\;x_t\mapsto \hat{x}_0\in\mathbb{R}^n$:
\begin{equation}
\hat{x}_0 \;=\; R_t(x_t;\theta).
\label{eq:reverse}
\end{equation}
We index timesteps either by an integer $t$ or by a ratio $t_{\!\mathrm{ratio}}\in(0,1]$, mapped as
$t=\lfloor t_{\!\mathrm{ratio}}(T{-}1)\rfloor{+}1$.

\subsection{Time-normalized latent perturbation}
\label{ssec:tnp}
Perturbations at different $t$ are not directly comparable because the \emph{forward} noise variance
is $1{-}\bar\alpha_t$. We therefore inject a \emph{time-normalized} perturbation at $x_t$:
\begin{equation}
\delta_t \;=\; \sigma_t\,\tilde{\delta},
\qquad 
\sigma_t=\sqrt{1-\bar\alpha_t},
\quad
\|\tilde{\delta}\|_{p}\le\eta,
\label{eq:delta}
\end{equation}
with $p\!\in\!\{2,\infty\}$ and a global budget $\eta{>}0$. The auxiliary variable
$\tilde{\delta}$ is drawn i.i.d.\ per sample from a \emph{Gaussian} direction and is the optimization
variable in our attack.\footnote{Rademacher directions gave similar trends in pilot runs; we fix
Gaussian for reproducibility.} The choice $\delta_t{=}\sqrt{1-\bar\alpha_t}\,\tilde\delta$ matches the
forward variance, equalizing the signal-to-noise ratio across timesteps and making budgets
comparable. For each pair $(x_0,t)$, we \emph{fix} the forward $\epsilon$ via a sample-specific
seed so that paired evaluations isolate the effect of $\delta_t$; across repetitions we average
over $K$ seeds. A second reverse pass from the perturbed latent yields
\begin{equation}
\hat{x}_0^{\delta} \;=\; R_t\!\big(x_t+\delta_t;\theta\big).
\label{eq:reverse_pert}
\end{equation}

We quantify degradation by a differentiable distance $D(\cdot,\cdot)$ computed on waveforms. Unless otherwise stated, we use the following metrics: \emph{CDPAM}, multi-resolution STFT (MR-STFT) distance, log-mel MSE, and waveform MSE. Gradients of the \emph{primary} metric are used \emph{only} inside the attack optimization; all metrics are reported at inference for robustness. The inner-loop PGD and its interaction with $R_t$ and $D$ are depicted in Fig.~\ref{fig:overview}(b).

\subsection{Objectives and threshold calibration}
\label{ssec:objectives}
We study two complementary objectives. For brevity, let
\begin{equation}
\hat{x}_0 \;=\; R_t(x_t;\theta),\qquad
\hat{x}_0^{\tilde\delta} \;=\; R_t\!\big(x_t+\sigma_t\tilde{\delta};\theta\big).
\label{eq:shorthand}
\end{equation}

\noindent\textbf{(O1) Fixed-budget maximal degradation}
\begin{equation}
\label{eq:fixed}
\max_{\;\|\tilde{\delta}\|_{p}\le \eta}\;\; D\!\big(\hat{x}_0,\;\hat{x}_0^{\tilde\delta}\big).
\end{equation}

\noindent\textbf{(O2) Adversarial cost}
\begin{equation}
\label{eq:cadv}
C_{\mathrm{adv}}(x_0;t,\tau)\;=\;
\inf\!\left\{\eta\!\ge\!0\ \middle|\ \exists\,\tilde{\delta},\;
\|\tilde{\delta}\|_{p}\!\le\!\eta,\;
D\!\big(\hat{x}_0,\hat{x}_0^{\tilde\delta}\big)\!\ge\!\tau\right\}.
\end{equation}

\noindent\textit{Calibration of $\tau$.}
We pre-register $\tau$ on a disjoint, unlabeled \emph{development non-member} set
$\mathcal{D}_\mathrm{dev}$. For each $(x_0,t)\!\in\!\mathcal{D}_\mathrm{dev}$ and $L$ random unit
Gaussian directions $u$, we evaluate
\begin{equation}
\label{eq:taucal}
D\!\Big(\hat{x}_0,\; R_t\!\big(x_t+\sigma_t\,\eta_{\mathrm{ref}}\,u;\theta\big)\Big),
\end{equation}
with the forward $\epsilon$ fixed per $(x_0,t)$. We set $\tau$ to the 95th percentile over all samples and directions (default $\eta_{\mathrm{ref}}{=}0.05$). We fix $\tau$ to P95 throughout experiments. The outer budget search yielding $C_{\mathrm{adv}}$ and the subsequent membership decision are illustrated in Fig.~\ref{fig:overview}(a).

\subsection{Optimization and compute parity}
\label{ssec:opt}
We solve (O1) by projected gradient descent (PGD) on $\tilde{\delta}$; (O2) uses an outer
\emph{bisection} on $\eta\!\in\![0,\eta_{\max}]$ with inner PGD. Inner PGD employs momentum $0.9$,
restarts $r$, and a step size tied to the current budget,
\[
\alpha \;=\; \beta\,\frac{\eta}{K},\quad \beta\in[0.2,0.3],
\]
with $K$ steps. We project by
$\Pi_{\|\cdot\|_2\le\eta}(\mathbf{z})=\eta\,\mathbf{z}/\max(\eta,\|\mathbf{z}\|_2)$ for $p{=}2$, and
elementwise clipping for $p{=}\infty$. Early stopping triggers when
$\Delta D/D\!<\!1\%$ for $3$ consecutive steps or $\|\nabla\|_2\!<\!10^{-6}$. Outer bisection runs
$10$ steps. Gradients flow through $R_t$ and the chosen $D$; we enable gradient checkpointing
across reverse steps. To ensure fair comparisons, we report UNet calls (forward/backward), wall-clock, and estimated FLOPs. With PGD($K$) and bisection($B$) the attack uses approximately $(K{+}2)B$ reverse passes and $(K{+}1)B$ metric evaluations per sample; baselines are repeated or multi-$t$ to match total compute within $\pm 5\%$.

\begin{algorithm}[t]
\caption{Adversarial cost $C_{\mathrm{adv}}$ at timestep $t$ (DDIM, $p\!\in\!\{2,\infty\}$)}
\label{alg:cadv}
\begin{algorithmic}[1]
\Require $x_0$, schedule $\{\bar\alpha_s\}$, $t$, reverse $R_t$, metric $D$, norm $p$, max budget $\eta_{\max}$, steps $K$, restarts $r$, step-scale $\beta$, seed $s$
\State Fix forward $\epsilon$ by seed $s$ and form $x_t$ via~\eqref{eq:forward}; set $\sigma_t \gets \sqrt{1-\bar\alpha_t}$
\State Initialize $[l,u] \gets [0,\eta_{\max}]$
\For{$b \gets 1$ to $10$} \Comment{bisection (Fig.~\ref{fig:overview}a)}
  \State $\eta \gets (l{+}u)/2$; \quad $D^\star \gets 0$
  \For{$j \gets 1$ to $r$} \Comment{PGD inner loop (Fig.~\ref{fig:overview}b)}
    \State Sample $\tilde\delta^{(0)} \sim \mathcal{N}(0,I)$ and project to $\lVert\cdot\rVert_p \le \eta$
    \For{$k \gets 0$ to $K{-}1$}
      \State $\hat{x}_0 \gets R_t(x_t)$; \quad $\hat{x}_0^\delta \gets R_t(x_t+\sigma_t \tilde\delta^{(k)})$
      \State $g \gets \nabla_{\tilde\delta}\, D(\hat{x}_0,\hat{x}_0^\delta)$; \quad update momentum
      \State $\tilde\delta^{(k{+}1)} \gets \Pi_{\lVert\cdot\rVert_p \le \eta}\!\bigl(\tilde\delta^{(k)} + \alpha\,\mathrm{step}(g)\bigr)$, \quad $\alpha \gets \beta\,\eta/K$
    \EndFor
    \State $D^\star \gets \max\!\bigl\{D^\star,\, D\bigl(R_t(x_t),\, R_t(x_t{+}\sigma_t\tilde\delta^{(K)})\bigr)\bigr\}$
  \EndFor
  \If{$D^\star \ge \tau$}
    \State $u \gets \eta$
  \Else
    \State $l \gets \eta$
  \EndIf
\EndFor
\State \Return $C_{\mathrm{adv}} \approx u$
\end{algorithmic}
\end{algorithm}

\subsection{Relation to trajectory-based MIAs}
Trajectory reconstruction attacks (e.g., proximal-initialization style) recover a ``ground-truth'' DDIM trajectory point $x_{t-t'}$ from $(x_0,x_t)$ and compare it with a model-predicted $x'_{t-t'}$ using an $\ell_p$ distance. Our LSA-Probe instead injects a \emph{time-normalized} perturbation at $x_t$ and searches a worst-case direction via PGD. The two families are complementary under a white-box threat model; we re-implement the trajectory method~\cite{kong2024an} as a baseline for head-to-head comparison under matched compute.

\subsection{Applicability to latent diffusion (MusicLDM)}
\label{ssec:ldm_note}
The probe applies unchanged to latent diffusion. In MusicLDM, diffusion operates on VAE latents $z$. For each waveform $x_0$, we compute $z_0=\mathrm{Enc}(x_0)$, forward-noise to $z_t$, and inject the time-normalized perturbation in latent space: $z_t \mapsto z_t + \delta_t$. The deterministic reverse $R_t(\cdot;\theta)$ runs in latent space; the waveform scored by $D$ is obtained with a \emph{frozen} decoder:
\begin{equation}
\hat{x}_0=\mathrm{Dec}\!\big(R_t(z_t;\theta)\big),\qquad
\hat{x}_0^{\tilde\delta}=\mathrm{Dec}\!\big(R_t(z_t+\sigma_t\tilde\delta;\theta)\big).
\end{equation}
Gradients backpropagate through $R_t$ and the fixed decoder; when conditioning is used, the text prompt and guidance are held constant across paired evaluations. Compute parity counts both UNet and decoder calls.

\begin{table*}[t]
  \centering
  \small
  \setlength{\tabcolsep}{6pt}
  \caption{Main results under matched compute (DDIM, $t_{\mathrm{ratio}}{=}0.6$, $p{=}2$, $\eta_{\max}{=}0.8$).
  We report TPR@\SI{1}{\percent} FPR and AUC-ROC (mean$\pm$95\% CI).
  ``Best Baseline'' is the best among Loss / Trajectory / SecMI.
  $\Delta$ is Ours minus Best Baseline.}
  \label{tab:main_compact}
  \begin{tabular}{llcccc}
    \toprule
    \multicolumn{1}{l}{Model} & \multicolumn{1}{l}{Dataset}
    & \multicolumn{1}{c}{Best Baseline (TPR@1\% / AUC)}
    & \multicolumn{1}{c}{Ours (TPR@1\% / AUC)}
    & \multicolumn{2}{c}{$\Delta$ (Ours $-$ Best)} \\
    \cmidrule(lr){3-3}\cmidrule(lr){4-4}\cmidrule(lr){5-6}
    & & \multicolumn{1}{c}{--} & \multicolumn{1}{c}{--} & \multicolumn{1}{c}{TPR@1\%} & \multicolumn{1}{c}{AUC} \\
    \midrule
    MusicLDM & MAESTRO
      & $0.10\ \bigl(0.07\text{--}0.12\bigr) / 0.58 \pm 0.02$
      & $\mathbf{0.13}\ \bigl(\mathbf{0.10}\text{--}\mathbf{0.15}\bigr) / \mathbf{0.61}\pm\mathbf{0.03}$
      & $+0.03$ & $+0.03$ \\
    MusicLDM & FMA-Large
      & $0.08\ \bigl(0.05\text{--}0.10\bigr) / 0.56 \pm 0.01$
      & $\mathbf{0.14}\ \bigl(\mathbf{0.10}\text{--}\mathbf{0.16}\bigr) / \mathbf{0.59}\pm\mathbf{0.02}$
      & $+0.06$ & $+0.03$ \\
    DiffWave & MAESTRO
      & $0.12\ \bigl(0.09\text{--}0.15\bigr) / 0.63 \pm 0.02$
      & $\mathbf{0.20}\ \bigl(\mathbf{0.16}\text{--}\mathbf{0.24}\bigr) / \mathbf{0.67}\pm\mathbf{0.02}$
      & $+0.08$ & $+0.04$ \\
    DiffWave & FMA-Large
      & $0.11\ \bigl(0.08\text{--}0.14\bigr) / 0.62 \pm 0.02$
      & $\mathbf{0.18}\ \bigl(\mathbf{0.14}\text{--}\mathbf{0.22}\bigr) / \mathbf{0.66}\pm\mathbf{0.02}$
      & $+0.07$ & $+0.04$ \\
    \bottomrule
  \end{tabular}
\end{table*}

\begin{figure*}[t]
  \centering
  \begin{subfigure}[t]{0.245\textwidth}
    \centering
    \IfFileExists{fig_a_roc.pdf}{
      \includegraphics[width=\linewidth]{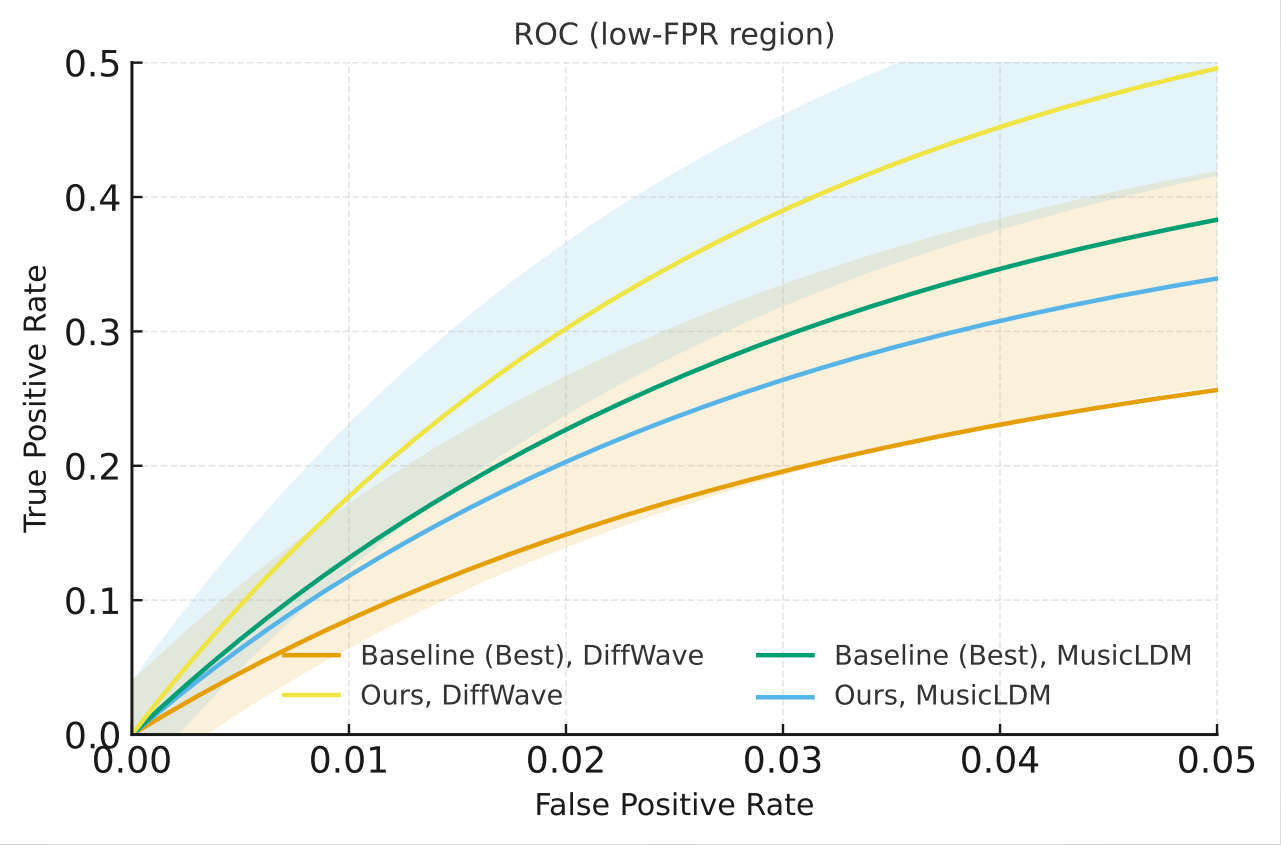}
    }{
      \fbox{\rule{0pt}{33mm}\rule{0.98\linewidth}{0pt}}
    }
    \caption{ROC @ $t_{\mathrm{ratio}}{=}0.6$ (DiffWave / MusicLDM).}
    \label{fig:abl4_roc}
  \end{subfigure}
  \hfill
  \begin{subfigure}[t]{0.245\textwidth}
    \centering
    \IfFileExists{fig_b_timestep.pdf}{
      \includegraphics[width=\linewidth]{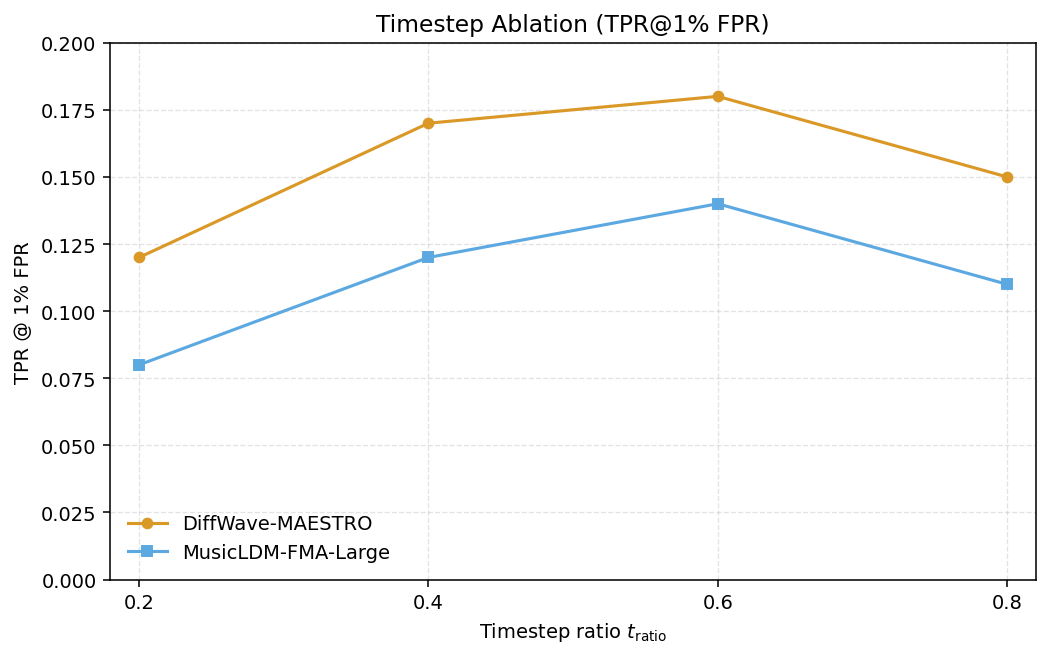}
    }{
      \fbox{\rule{0pt}{33mm}\rule{0.98\linewidth}{0pt}}
    }
    \caption{$t_{\mathrm{ratio}}\in\{0.2,0.4,0.6,0.8\}$.}
    \label{fig:abl4_t}
  \end{subfigure}
  \hfill
  \begin{subfigure}[t]{0.245\textwidth}
    \centering
    \IfFileExists{fig_c_budget.pdf}{
      \includegraphics[width=\linewidth]{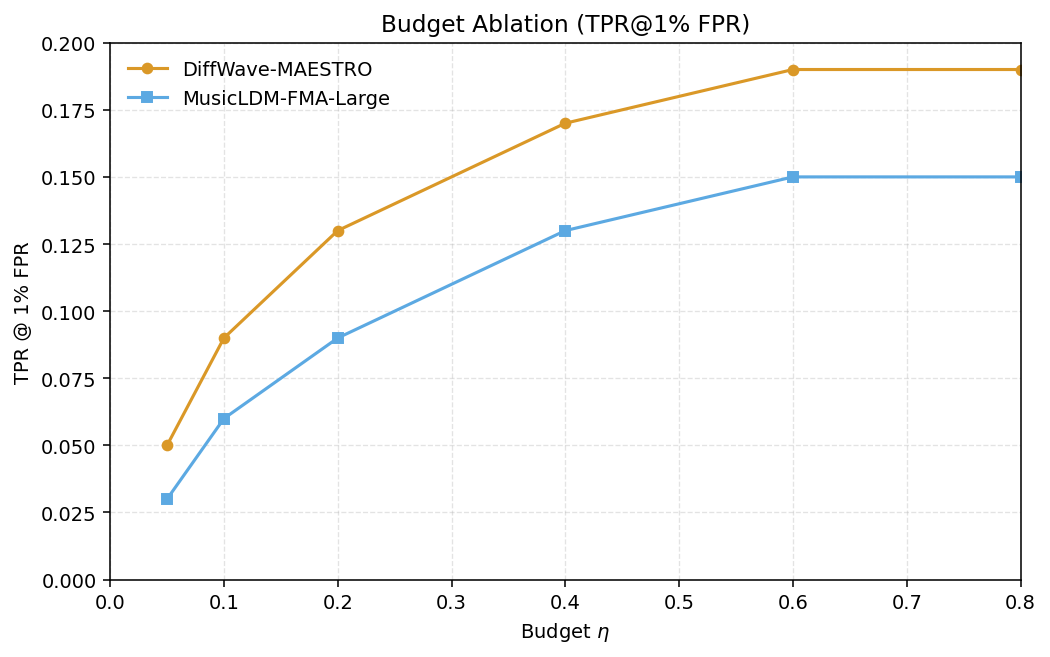}
    }{
      \fbox{\rule{0pt}{33mm}\rule{0.98\linewidth}{0pt}}
    }
    \caption{Budget $\eta\in[0.05,0.8]$.}
    \label{fig:abl4_eta}
  \end{subfigure}
  \hfill
  \begin{subfigure}[t]{0.245\textwidth}
    \centering
    \IfFileExists{fig_d_metric.pdf}{
      \includegraphics[width=\linewidth]{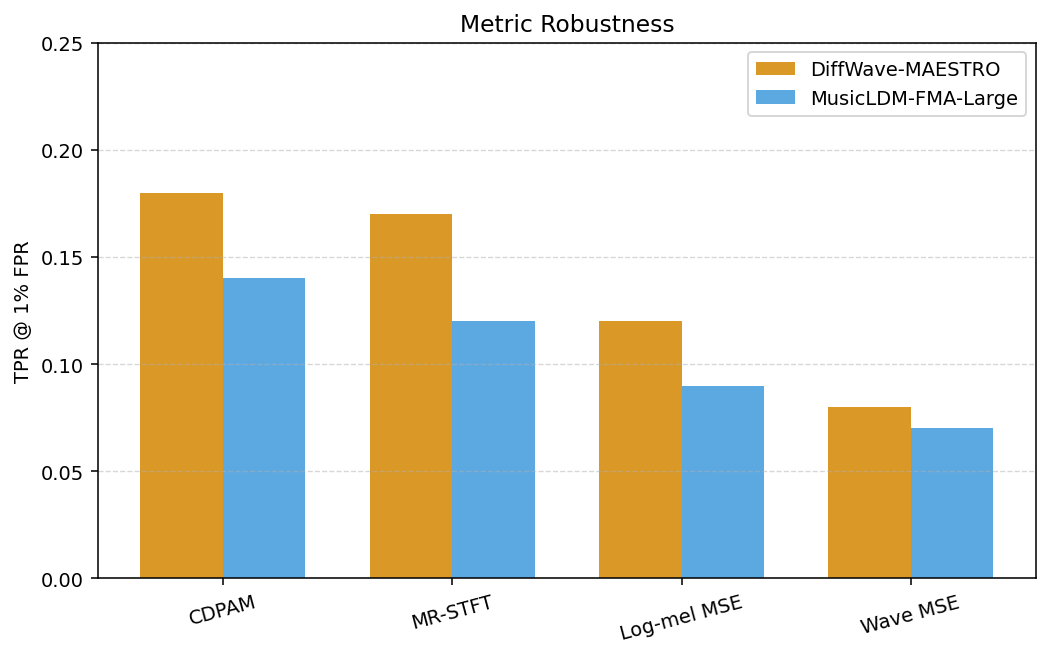}
    }{
      \fbox{\rule{0pt}{33mm}\rule{0.98\linewidth}{0pt}}
    }
    \caption{$D\in\{$CDPAM, MR-STFT, log-mel MSE, wave MSE$\}$.}
    \label{fig:abl4_metric}
  \end{subfigure}
  \vspace{-1mm}
  \caption{\textbf{Key analyses under fixed $\tau{=}$P95} (DDIM, $p{=}2$, $\eta_{\max}{=}0.8$). 
  (a) ROC curves with 95\% CIs; our LSA-Probe improves the low-FPR region. 
  (b) Timestep ablation: mid-trajectory timesteps yield stronger separability. 
  (c) Budget ablation: larger budgets help until mild saturation. 
  (d) Metric robustness: perceptual metrics (CDPAM/MR-STFT) outperform training-aligned MSEs at low FPR.}
  \label{fig:ablations4}
\end{figure*}

\section{Experiments}
\label{sec:experiments}

We evaluate two music diffusion families: (i) a waveform DDPM instantiated as \textbf{DiffWave}~\cite{kong2020diffwave} and (ii) a latent diffusion model instantiated as \textbf{MusicLDM} (unconditional/weakly conditioned LDM operating in VAE latents)~\cite{chen2024musicldm}. Datasets are \textbf{MAESTRO v3} (solo piano)~\cite{hawthorne2019maestro} and \textbf{FMA-Large} (multi-genre)~\cite{defferrard2017fma}. We adopt deterministic DDIM sampling, timestep ratio $t_{\mathrm{ratio}}{=}0.6$, norm $p{=}2$, maximum budget $\eta_{\max}{=}0.8$, and the primary distance $D$ is CDPAM; MR-STFT, log-mel MSE, and waveform MSE are also reported for robustness. Audio is trained at 22.05\,kHz and metrics are computed at 16\,kHz via in-graph resampling. Our threat model is white-box, developer-side with gradient access. According to previous research~\cite{Carlini2023DiffusionMIA, ART_LabelOnly_Calibration, LiZhang2021LabelOnly}, the decision threshold is fixed at $\tau=\mathrm{P95}$ on a disjoint development non-member set.

\vspace{-1mm}
\subsection{Evaluation metrics and protocol}
We report AUC-ROC, TPR@FPR=\SI{1}{\percent}, TPR@FPR=\SI{0.1}{\percent}. AUC confidence intervals are computed with \emph{DeLong}~\cite{DeLong1988}; other endpoints use \emph{sample-level bootstrap} (10{,}000 resamples). Across timesteps and metrics, we control family-wise error via Holm--Bonferroni~\cite{Holm1979}. For membership labels, the \emph{member} pool consists of training clips; the \emph{non-member} pool is sampled from held-out validation/test with near-/cross-domain controls. We remove duplicates/covers using Chromaprint+LSH with manual spot checks.

\subsection{Models, training, and splits}
\label{subsec:train}
DiffWave follows the official UNet architecture (200k steps, batch 32, AdamW, cosine LR). MusicLDM is trained unconditionally in the VAE latent space (stride $\times4$); during attacks, the decoder is frozen and gradients flow through the latent reverse operator while distances are computed on decoded waveforms. We cut audio into 4\,s clips, stratified by piece/artist (MAESTRO) or track/artist (FMA) to avoid leakage. Unless otherwise specified, samplers are deterministic (DDIM, $\sigma_t{=}0$).

\subsection{Baselines and compute accounting}
Baselines include: loss-aligned (denoising/reconstruction losses at $t$ and at the endpoint)~\cite{pang2025white}, trajectory (PIA/PIAN)~\cite{kong2024an}, SecMI~\cite{li2025towards}. To ensure fairness, we \emph{match compute within $\pm5\%$} using per-sample wall-clock on A100-80GB. Our adversarial cost $C_{\mathrm{adv}}$ uses outer bisection ($B{=}10$) with inner PGD ($K{=}12$ steps, restarts $r{=}5$), which totals approximately $(K{+}2)B$ reverse passes per sample.

\subsection{Main results}
\label{subsec:main}
Table~\ref{tab:main_compact} reports the primary endpoint (TPR@\SI{1}{\percent}FPR) and AUC-ROC under matched compute (DDIM, $t_{\mathrm{ratio}}{=}0.6$, $p{=}2$, $\eta_{\max}{=}0.8$). Across both datasets and model families, LSA-Probe consistently improves the low-FPR operating point over the strongest baseline: for MusicLDM (LDM), TPR@\SI{1}{\percent} increases from $0.10\!\rightarrow\!0.13$ on MAESTRO and $0.08\!\rightarrow\!0.14$ on FMA-Large (absolute gains $+0.03$ and $+0.06$), with AUC lifts of $+0.03$ on both datasets; for DiffWave (DDPM), TPR@\SI{1}{\percent} rises from $0.12\!\rightarrow\!0.20$ (MAESTRO) and $0.11\!\rightarrow\!0.18$ (FMA-Large) (gains $+0.08$ and $+0.07$), accompanied by larger AUC improvements of $+0.04$ on both datasets. Importantly, even with these gains, LDMs remain more robust to MIA than DDPMs.

\subsection{Key analyses: ROC, timestep, budget, and metric}
Figure~\ref{fig:ablations4} consolidates the four most informative analyses into a single 4$\times$1 panel. 
\textbf{(a) ROC:} At $t_{\mathrm{ratio}}{=}0.6$, LSA-Probe improves the low-FPR region across models, consistent with Table~\ref{tab:main_compact}. 
\textbf{(b) Timestep:} Sweeping $t_{\mathrm{ratio}}\!\in\!\{0.2,0.4,0.6,0.8\}$ shows peak separability at mid-trajectory (0.6), aligning with the intuition that the reverse path transitions from coarse global layout to finer details. 
\textbf{(c) Budget:} Increasing $\eta$ improves TPR@1\% until mild saturation near $\eta{\approx}0.6$--$0.8$, where saturation rates remain low. 
\textbf{(d) Metric:} Perceptual metrics (CDPAM, MR-STFT) provide stronger discrimination at low FPR than training-aligned MSEs, while the latter remain useful complementary views.

\subsection{Implementation details}
We fix the forward noise per $(x_0,t)$ using a sample-specific seed so that paired evaluations isolate the effect of $\delta_t$; across repetitions we average over $S$ random seeds. The adversarial inner loop uses PGD with $K{=}12$ steps, momentum $0.9$, restarts $r{=}5$, and step size $\alpha=\beta\,\eta/K$ with $\beta\!\in[0.2,0.3]$; Early stopping triggers at $\Delta D/D{<}1\%$ for three consecutive steps or $\|\nabla\|_2{<}10^{-6}$.

\section{Conclusion}
\label{sec:conclusion}

We proposed \textbf{LSA-Probe}, a white-box membership inference attack for music diffusion models that measures
time-normalized latent stability along the reverse trajectory, using the \emph{adversarial budget} required to exceed a fixed perceptual degradation threshold as the score.
Across two model families (DiffWave and MusicLDM) and two corpora (MAESTRO v3, FMA-Large), LSA-Probe consistently improves low-FPR detection over strong baselines under matched compute.
Ablations indicate that mid-trajectory timesteps and moderate budgets are most informative, and that perceptual metrics
(CDPAM/MR-STFT) reveal stability gaps more reliably than MSE-based distances.

\section{Acknowledgements}
This work was supported by Basic Research Program of JiangSu Province under Grant BG2024027, and the XJTLU Research Development Fund (Grant No. RDF-22-02-046).

\bibliographystyle{IEEEbib}
\bibliography{strings,refs}

@inproceedings{chen2024musicldm,
  title={{MusicLDM}: Enhancing Novelty in Text-to-Music Generation Using Beat-Synchronous Mixup Strategies},
  author={Chen, Ke and Wu, Yusong and Liu, Haohe and Nezhurina, Marianna and Berg-Kirkpatrick, Taylor and Dubnov, Shlomo},
  booktitle={ICASSP 2024 - 2024 IEEE International Conference on Acoustics, Speech and Signal Processing (ICASSP)},
  year={2024},
  organization={IEEE}
}

@inproceedings{defferrard2017fma,
  title={{FMA}: A Dataset For Music Analysis},
  author={Defferrard, Micha{\"e}l and Benzi, Kirell and Vandergheynst, Pierre and Bresson, Xavier},
  booktitle={18th International Society for Music Information Retrieval Conference (ISMIR)},
  year={2017},
  pages={316--323}
}

@article{dornis2024generative,
  title={Generative {AI} Training and Copyright Law},
  author={Dornis, Tim W and Stober, Sebastian},
  journal={arXiv preprint arXiv:2502.15858},
  year={2024}
}

@inproceedings{hawthorne2019maestro,
  title={Enabling Factorized Piano Music Modeling and Generation with the {MAESTRO} Dataset},
  author={Hawthorne, Curtis and Stasyuk, Andriy and Roberts, Adam and Simon, Ian and Fu, Cheng-Zhi and Green, Sanders and Dieleman, Sander and Hentschel, Erich and Eck, Douglas},
  booktitle={International Conference on Learning Representations (ICLR)},
  year={2019}
}

@inproceedings{hilprecht2019monte,
  title={Monte Carlo and Reconstruction Membership Inference Attacks against Generative Models},
  author={Hilprecht, Benjamin and H{\"a}rterich, Martin and Bernau, Daniel},
  booktitle={Proceedings on Privacy Enhancing Technologies},
  volume={2019},
  number={4},
  pages={232--249},
  year={2019},
  organization={Sciendo}
}

@inproceedings{ho2020denoising,
  title={Denoising Diffusion Probabilistic Models},
  author={Ho, Jonathan and Jain, Ajay and Abbeel, Pieter},
  booktitle={Advances in Neural Information Processing Systems},
  volume={33},
  pages={6840--6851},
  year={2020}
}

@inproceedings{keskar2017on,
  title={On Large-Batch Training for Deep Learning: Generalization Gap and Sharp Minima},
  author={Keskar, Nitish Shirish and Mudigere, Dheevatsa and Nocedal, Jorge and Smelyanskiy, Mikhail and Tang, Ping Tak Peter},
  booktitle={International Conference on Learning Representations (ICLR)},
  year={2017}
}

@inproceedings{kong2020diffwave,
  title={{DiffWave}: A Versatile Diffusion Model for Audio Synthesis},
  author={Kong, Zhifeng and Ping, Wei and Huang, Jiaji and Zhao, Kexin and Catanzaro, Bryan},
  booktitle={International Conference on Learning Representations (ICLR)},
  year={2021}
}

@inproceedings{chowdhury2024melfusion,
  title={Melfusion: Synthesizing music from image and language cues using diffusion models},
  author={Chowdhury, Sanjoy and Nag, Sayan and Joseph, KJ and Srinivasan, Balaji Vasan and Manocha, Dinesh},
  booktitle={Proceedings of the IEEE/CVF Conference on Computer Vision and Pattern Recognition},
  pages={26826--26835},
  year={2024}
}

@inproceedings{matsumoto2023membership,
  title={Membership Inference Attacks against Diffusion Models},
  author={Matsumoto, Tomoya and Miura, Takayuki and Yanai, Naoto},
  booktitle={2023 IEEE Security and Privacy Workshops (SPW)},
  pages={310--321},
  year={2023},
  organization={IEEE}
}

@article{rossi2025membership,
  title={Membership Inference Attacks on Sequence Models},
  author={Rossi, Lorenzo and Aerni, Michael and Zhang, Jie and Tram{\`e}r, Florian},
  journal={arXiv preprint arXiv:2506.05126},
  year={2025}
}

@inproceedings{shokri2017membership,
  title={Membership Inference Attacks Against Machine Learning Models},
  author={Shokri, Reza and Stronati, Marco and Song, Congzheng and Shmatikov, Vitaly},
  booktitle={2017 IEEE Symposium on Security and Privacy (SP)},
  pages={3--18},
  year={2017},
  organization={IEEE}
}

@article{tirumala2022memorization,
  title={Memorization without overfitting: Analyzing the training dynamics of large language models},
  author={Tirumala, Kushal and Markosyan, Aram and Zettlemoyer, Luke and Aghajanyan, Armen},
  journal={Advances in Neural Information Processing Systems},
  volume={35},
  pages={38274--38290},
  year={2022}
}

@inproceedings{
kong2024an,
title={An Efficient Membership Inference Attack for the Diffusion Model by Proximal Initialization},
author={Fei Kong and Jinhao Duan and RuiPeng Ma and Heng Tao Shen and Xiaoshuang Shi and Xiaofeng Zhu and Kaidi Xu},
booktitle={The Twelfth International Conference on Learning Representations (ICLR)},
year={2024},
url={https://openreview.net/forum?id=rpH9FcCEV6}
}

@inproceedings{liu2022membership,
  title={Membership inference attacks by exploiting loss trajectory},
  author={Liu, Yiyong and Zhao, Zhengyu and Backes, Michael and Zhang, Yang},
  booktitle={Proceedings of the 2022 ACM SIGSAC Conference on Computer and Communications Security},
  pages={2085--2098},
  year={2022}
}

@article{DeLong1988,
  author  = {DeLong, Elizabeth R. and DeLong, David M. and Clarke-Pearson, Daniel L.},
  title   = {Comparing the Areas under Two or More Correlated Receiver Operating Characteristic Curves: A Nonparametric Approach},
  journal = {Biometrics},
  year    = {1988},
  volume  = {44},
  number  = {3},
  pages   = {837--845},
  doi     = {10.2307/2531595}
}

@article{Holm1979,
  author  = {Holm, Sture},
  title   = {A Simple Sequentially Rejective Multiple Test Procedure},
  journal = {Scandinavian Journal of Statistics},
  year    = {1979},
  volume  = {6},
  number  = {2},
  pages   = {65--70}
}

@inproceedings{li2025towards,
  title={Towards Black-Box Membership Inference Attack for Diffusion Models},
  author={Li, Jingwei and Dong, Jing and He, Tianxing and Zhang, Jingzhao},
  booktitle={ICLR 2025 Workshop on Deep Generative Model in Machine Learning: Theory, Principle and Efficacy}
}

@article{pang2025white,
  title={White-box Membership Inference Attacks against Diffusion Models},
  author={Pang, Yan and Wang, Tianhao and Kang, Xuhui and Huai, Mengdi and Zhang, Yang},
  journal={Proceedings on Privacy Enhancing Technologies},
  year={2025}
}

@inproceedings{LiZhang2021LabelOnly,
  author    = {Zheng Li and Yang Zhang},
  title     = {Membership Leakage in Label-Only Exposures of {ML} Models},
  booktitle = {Proceedings of the 2021 ACM SIGSAC Conference on Computer and Communications Security (CCS)},
  year      = {2021},
  }

@misc{ART_LabelOnly_Calibration,
  author       = {{Adversarial Robustness Toolbox Developers}},
  title        = {Label-only membership inference: calibrate distance threshold using top-$t$ percentile},
  howpublished = {\url{https://adversarial-robustness-toolbox.readthedocs.io/en/latest/modules/attacks/inference/membership_inference.html}},
  year         = {2024},

}

@inproceedings{Carlini2023DiffusionMIA,
  author    = {Nicholas Carlini and others},
  title     = {Extracting Training Data from Diffusion Models},
  booktitle = {USENIX Security},
  year      = {2023},
  }

\end{document}